\documentclass[12pt,a4,fleqn,onecolumn,oneside]{article} 
\usepackage{amsmath,amsfonts,graphicx,amsbsy}
\usepackage[amssymb]{SIunits}
\usepackage{lineno}
\begin{document}
\title{ \bf Measurement of cluster  elongation and charge  in a  pixel detector of 10~$\mu$m pitch at sub-GeV energies}
\author{\bf M.~Adamus$^1$, J.~Ciborowski$^2$, {\L}.~M\c{a}czewski$^2$, P.~{\L}u\.{z}niak$^3$}
\date{\today}
\maketitle

\vspace{1cm}

\begin{center}
$^1$ National Centre for Nuclear Studies, Ho\.{z}a 69, 00-681 Warszawa, Poland,\\

$^2$ University of Warsaw, Faculty of Physics, Ho\.{z}a 69, 00-681 Warszawa, Poland, \\

$^3$ University of {\L}\'{o}d\'{z}, Faculty of  Physics and Applied Informatics,\\ Pomorska 149/153,  90-236 {\L}\'{o}d\'{z}, Poland,

\end{center}

\noindent
\small { {\bf Abstract:}
 We present measurements of elongation and cluster charge using   MIMOSA-18  MAPS pixel matrix  with  10~$\mu$m pixel pitch, using electron test beams of energies ranging from 15 to 500 MeV. We observe energy dependence of cluster  charge and elongation for large incident angles.

 }

\vspace{2cm}

\noindent
In our recent publications~\cite{cite:1stpaper},\cite{LMaczawski} we presented results of our studies of cluster shapes in a Monolithic Active Pixel Sensor  (MAPS)  MIMOSA-5 of  17~$\mu$m pixel pitch. The  MAPS pixel arrays have been  considered, among others, as a technology for  a vertex detector at the future  International Linear Collider~\cite{Accelerator},\cite{Detector}.
The vertex detector will be exposed
to a significant background of $e^+e^-$ pairs from the beamstrahlung process.
These electron-positron pairs enter the pixel matrices
and create clusters additional to those due to  secondary particles from  beam collisions~\cite{GuineaPig}, \cite{PLuzniak}.
A MIMOSA-5 detector was exposed to 1 and 6~GeV beams of electrons incident at various angles $\theta$.  The aim of these studies was to  measure  magnitude  of cluster elongation  and precision  of its azimuthal angle determination  as  functions of the angle $\theta$.
In this  paper we present   measurements of elongation and cluster charge using  the MIMOSA-18  pixel matrix  with  10~$\mu$m pixel pitch, using electron test beams of  lower energies, as compared to previous measurements, and several  energy settings,  at MAMI (Mainzer Mikrotron at Johannes Gutenberg Universit\"{a}t Mainz) and  INFN Frascati.

\noindent
Monolithic Active Pixel Sensor, MAPS,  consists of three layers: a highly doped p-type substrate, a partially depleted p-type epitaxial layer
(sensitive volume)  and a p-well  where the collecting diodes of the n-type are implemented.
Charge carriers,  created  during  passage of an ionising particle, diffuse  inside the
epitaxial layer. Potential barriers, appearing on the layer  boundaries  due to different doping levels,  act to
confine electrons within the sensitive volume until they are collected by the diodes.
Studies show that the MAPS matrices may have spatial resolution of the order a few $\mu$m,
almost 100\% detection efficiency, possess good resistance to radiation~\cite{Deptuch1},\cite{MAPS1}.
We have used the  MIMOSA-18\footnote{\textbf{M}inimum \textbf{I}onizing particle \textbf{MOS} \textbf{A}ctive pixel sensor} MAPS prototype for the present studies. The device was fabricated in the AMS $0.35~\mu$m opto CMOS technology. The epitaxial layer, which constitutes the sensitive volume of the detector, was $14~\mu$m thick. We have used a module of $256\times 256$ pixels.
The MIMOSA-18 matrix was mounted on an adjustable mechanical support, enabling rotations, and was kept at room temperature.  The matrix was oriented  manually to the desired angles before each data taking run using the angular scale with an accuracy of approximately $\pm 1^\circ$.
The matrix was exposed to  electron  test beams as follows:
15~MeV at MAMI and 66, 100, 150, 200, 400, 500~MeV at INFN Frascati.
The measurements were done for various orientations of the matrix w.r.t. the beam.
The lowest energy (15~MeV) falls in the energy range of  beamstrahlung electrons in the  planned  ILC collider.
\begin{figure}
	\begin{center}
		\resizebox{8cm}{!}{
			\includegraphics[]{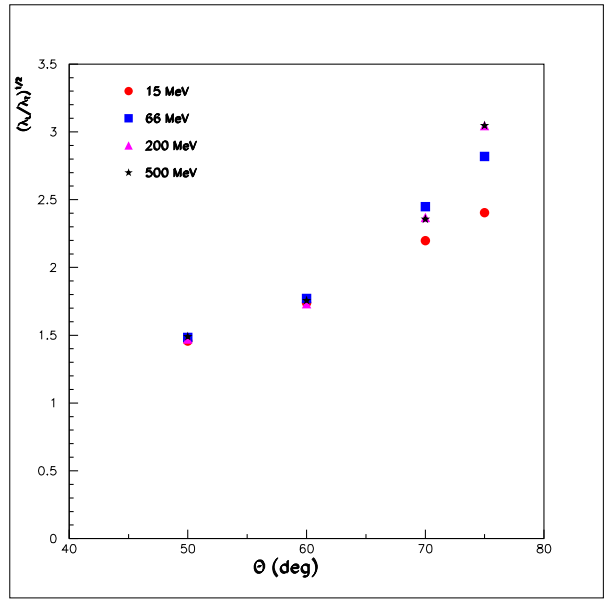}
		}
		\caption[]{Measured ratio of the longitudinal and transverse dimensions of a cluster as
a function of the $\theta$ angle for several beam energies.}
		\label{fig:wydluzenie}
	\end{center}
\end{figure}
Clusters of pixels were reconstructed according to the following procedure.
First,  pedestals and noise levels,  were evaluated for each pixel.
A cluster  was defined as a set of pixels, each with   charge  to noise ratio  grater than 3, with exception of the seed, i.e. the pixel carrying the largest charge, where this ratio was required to be greater than 8.
Pixels in a cluster had  at least one common edge  with other  pixels belonging to cluster  (one-pixel clusters were rejected).
It was assumed that the charge-weighted centre of gravity of a cluster coincided with the track position at the matrix surface.
Particles passing through the detector at low incident angles leave statistically round clusters.
Since in a MAPS detector the charge transport is by diffusion, it is expected that clusters arising from sufficiently inclined tracks should be elongated in the track direction projected on the pixel plane while unaltered in the perpendicular direction. We have measured the magnitude of the elongation as a function of the incident angle $\theta$  for various energies.

\noindent
Longitudinal and transverse dimensions of a cluster  may be obtained from diagonalising  the charge distribution matrix defined as:
\begin{equation}
\left(\begin{array}{cc}
	\sum_{i = 1}^{N_{c}}\frac{q_{i}}{Q}\left(x_{i}-\overline{x}\right)^{2} & \sum_{i = 1}^{N_{c}}\frac{q_{i}}{Q}\left(x_{i}-\overline{x}\right)\left(y_{i}-\overline{y}\right)\\
	\sum_{i = 1}^{N_{c}}\frac{q_{i}}{Q}\left(x_{i}-\overline{x}\right)\left(y_{i}-\overline{y}\right) & \sum_{i = 1}^{N_{c}}\frac{q_{i}}{Q}\left(y_{i}-\overline{y}\right)^{2} \\
	\end{array} \right),
\label{Matrix}
\end{equation}
where $Q$ is the cluster charge, $q_{i}$ is the charge of the $i$-th pixel, $x_{i}$, $y_{i}$ are its positions in the MIMOSA-18 coordinates system
and  $\overline{x}$,  $\overline{y}$ are the coordinates of the charge-weighted centre of gravity of the cluster:
\begin{equation}
\overline{x} = \sum_{i = 1}^{N_{c}}\frac{q_{i}}{Q} x_{i},~~~
\overline{y} = \sum_{i = 1}^{N_{c}}\frac{q_{i}}{Q} y_{i}.
\label{CentreOfGravity}
\end{equation}
Diagonalisation of~(\ref{Matrix}) allows to determine the eigenvectors and the respective eigenvalues, $\lambda_{L}$ and $\lambda_{T}$.
The ratio $\sqrt{\lambda_{L}/\lambda_{T}}$ is a measure of cluster elongation. The measured dependence of this quantity on the incident angle $\theta$ is shown in Fig.~\ref{fig:wydluzenie}. Cluster elongation depends strongly on $\theta$ for sufficiently large values. While no dependence  of  elongation on track energy was seen for GeV-energy  tracks~\cite{cite:1stpaper}, we do  observe this dependence for  energies in the range of the present measurement,  approximately for   $\theta > 60^{\circ}$. Tracks incident at low  angles do not form  elongated clusters  because the charge is symmetrically distributed around the seed.

\begin{figure}
	\begin{center}
		\resizebox{8cm}{!}{
			\includegraphics[]{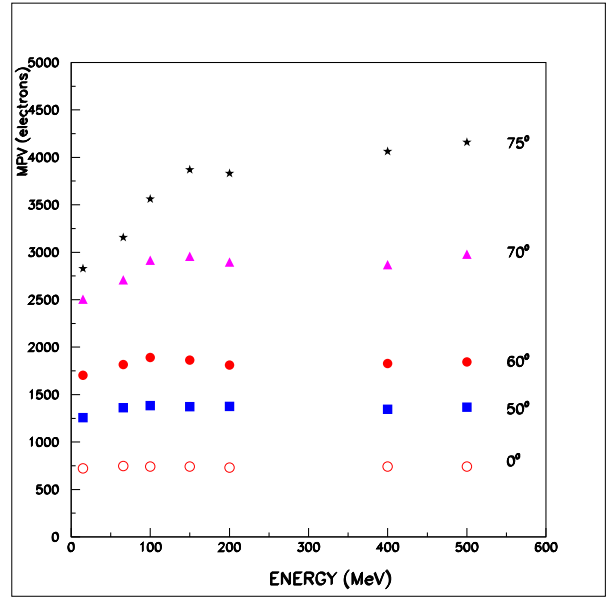}
		}
		\caption[]{MPV of the Landau charge distribution  as a function of beam energy for several incident angle settings, $\theta$.}
		\label{fig:vs_energy}
	\end{center}
\end{figure}

\begin{figure}
	\begin{center}
		\resizebox{8cm}{!}{
			\includegraphics[]{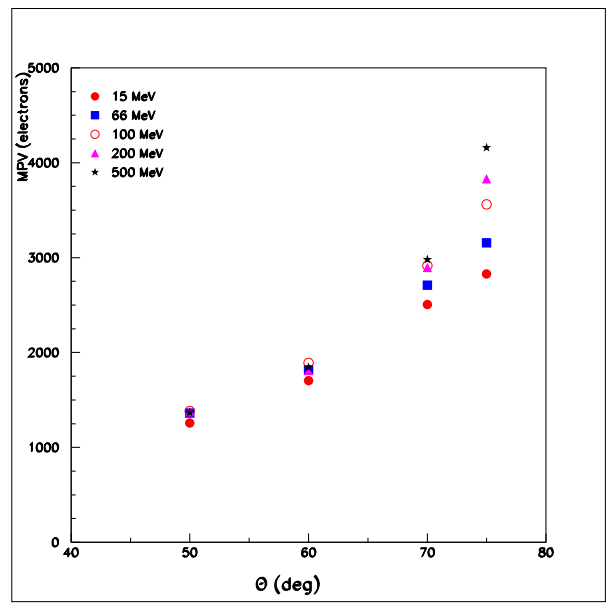}
		}
		\caption[]{MPV of the Landau charge distribution  as a function of incident angle, $\theta$, for several beam energies.}
		\label{fig:vs_theta}
	\end{center}
\end{figure}

\noindent
The charge of a  cluster is obtained by  summing  charges deposited in  pixels  belonging to it. The total  charge   is described by the Landau  distribution  with  the  most probable value, MPV,  depending  on  the track  energy and the $\theta$ angle.
Dependence of the MPV  on  track  energy  and  $\theta$ is shown in Figs.~\ref{fig:vs_energy} and  \ref{fig:vs_theta}, respectively.
We observe   that the larger $\theta$,  the stronger is the energy dependence of the MPV.
\noindent
 In summary, we have measured cluster shapes  and charges using the MIMOSA-18 MAPS detector exposed to 15~MeV at MAMI and 66, 100, 150, 200, 400, 500~MeV at INFN Frascati.
Cluster elongation was measured for various orientations of the pixel matrix w.r.t. the beam axis (angle $\theta$). The results may be summarised as follows: ($i$) clusters can be reliably reconstructed as elongated for  $\theta > 60^{\circ}$ and the effect grows rapidly with increasing $\theta$; ($ii$) energy dependence of the above effect is observed  for $\theta > 60^{\circ}$ where  elongation increases significantly with energy in the sub-GeV region (no such effect was present for 1~and 6~GeV tracks); ($iii$)  little energy dependence of the cluster charge  is observed for round clusters ($\theta < 60^{\circ}$)  while there is a significant energy dependence  for elongated clusters ($\theta > 60^{\circ}$);  clusters arising from beamstrahlung electron tracks carry  on average less charge that those from interaction secondaries.

\noindent
{\bf Acknowledgements}

\noindent
We thank the MAMI and  INFN Frascati Directorates  for the possibility to use the test beam facilities, the IReS (Strasbourg) and in particular Dr~M.~Winter for lending us the MIMOSA-18 matrix and Dr~M.~I.~Gregor  from DESY for her assistance in various fields. We thank  Dr K.~Aulenbacher from  Universit\"{a}t Mainz for his  help  in conducting the measurements, as well as  Dr G.~Mazzitelli and Dr B.~Buonomo for their assistance in Frascati.  This work was partially supported by the Commission of the European Communities under the 6$^{th}$ and 7$^{th}$ Framework Programmes "Structuring the European Research Area".


\begin{thebibliography}{00}

\bibitem{cite:1stpaper} {\L}.~M\c{a}czewski {\em et al.}, Nucl. Instrum. Meth. A610 (2009) 640;
e-Print: arXiv:0903.3658.
\bibitem{LMaczawski} {\L}.~M\c{a}czewski, Ph.D. thesis, University of Warsaw, Poland, 2010; e-Print: arXiv:1005.3710, 2010.
\bibitem{Accelerator} ILC Reference Design Report Volume 3 - Accelerator, arXiv:0712.2361, 2007.
\bibitem{Detector} ILC Reference Design Report Volume 4 - Detectors, arXiv:0712.2356, 2007.
\bibitem{GuineaPig} C. Rimbault {\em et al.}, "Study of incoherent pair generation in Guinea-Pig",
EUROTeV-Report-2005-016-1, 2005.
\bibitem{PLuzniak} P.~{\L}u\.zniak, in:  Proceedings of the International Linear Collider Workshop, Hamburg, 2007, edited by S. Riemann, eConf C0705302, Trk07 (2007); P.~{\L}u\.zniak, Ph.D. thesis, {\L}\'od\'z University, Poland, 2009.
\bibitem{Deptuch1} G. Deptuch {\em et al.}, Nucl. Instr. and Meth. Phys. Res. A 511 (2003) 240-249.
\bibitem{MAPS1} R. Turchetta {\em et al.}, Nucl. Instr. and Meth. Phys. Res. A 458 (2001) 677.
\bibitem{VME-DAQ} G. Claus {\em et al.}, IEEE Trans. Nucl. Sci. NS-52 (4) (2005).
\bibitem{DevisBeam} D. Contarato {\em et al.}, Nucl. Instr. and Meth. Phys. Res. A 565 (2006) 119-125.

\end{thebibliography}
\end{document}